\documentclass{birkjour}
 \theoremstyle{definition}
 
 \theoremstyle{remark}

 \numberwithin{equation}{section}

\begin{document}

%
%
%
%
%
%
%
%
%

\title[Dynamics of a Vacuum Bianchi Type V Universe]
 {Dynamics of a Vacuum Bianchi Type V Universe \\ with an Arbitrary Cosmological Constant}

\author[Ikjyot Singh Kohli]{Ikjyot Singh Kohli}

\address{%
York University \\
Department of Mathematics and Statistics \\
Toronto, Ontario, Canada}

\email{isk@mathstat.yorku.ca}



\date{February 13, 2017}

\begin{abstract}
In this paper, we describe the dynamics of a Bianchi Type V vacuum universe with an arbitrary cosmological constant. We begin by using an orthonormal frame approach to write Einstein's field equations as a coupled system of first-order ordinary differential equations. The equilibrium points of the resulting dynamical system are found to be expanding and contracting de Sitter universe solutions, a Minkowski spacetime solution, and static Anti-de Sitter universe solutions, which are characterized by a negative cosmological constant in addition to constant negative spatial curvature. While the expanding de Sitter universe solution is found to be asymptotically stable for $\Lambda > 0$, we also found that the AdS solution is stable for $\Lambda < 0$, as solutions form closed trajectories around it. Further, we show that the different equilibrium points are related to each other by a series of bifurcations of $\Lambda$.
\end{abstract}

\maketitle
\section{Introduction}
The Friedmann-Lema\^{i}tre-Robertson-Walker (FLRW) models are spatially homogeneous and isotropic, and are thus a very restricted class of cosmological models. On the contrary, spatially homogeneous, but \emph{anisotropic} models allow the investigation of much more general behaviour than the FLRW models. According to \cite{elliscosmo}, these Bianchi cosmologies can represent anisotropic modes, including rotation and global magnetic fields. They may also be good approximations in regions where there are inhomogeneities, but the spatial gradients are small. 

Of particular interest in this work is the dynamics of a Bianchi Type V cosmological model with an arbitrary cosmological constant. That is, we take the cosmological constant to be a free parameter in the Einstein field equations.  Our interest stems from the fact that the Bianchi Type V model contains $k=-1$ FLRW models as a special case (in the isotropic limit). The $k=-1$ FLRW models are interesting in their own right. A particular example is that of Anti-de Sitter space, which in the cosmological context, is the companion spacetime of constant spacetime curvature, but, with $R < 0$. In FLRW form, it can be represented by
\begin{equation}
ds^2 = -dt^2 + \cos^2 t \left[d \chi^2 + \sinh^2 \chi \left(d\theta^2 + \sin^2 \theta d \phi^2 \right)\right],
\end{equation}
which, of course, only covers a part of the spacetime. Further, this metric requires $\Lambda < 0$, which contradicts observations, but is fundamental in string theory because of the AdS/CFT correspondence. The interested reader is encouraged to see \cite{elliscosmo} and references therein for further information on these topics. 

In the literature thus far, a study of the dynamics of Bianchi Type V models using dynamical systems methodologies has been carried out in various contexts. In particular, Coley and Hervik \cite{2005CQGra..22..579C} studied the dynamics of tilted Bianchi V models. Goliath and Nilsson \cite{2000JMP....41.6906G} studied the late-time isotropization of two-fluid Bianchi type V models. Christodoulakis, Kofinas, and Zarikas \cite{2000PhLA..275..182C} examined a Bianchi Type V model filled with diffuse matter and its relation to the CMBR. Shogin and Hervik \cite{2015CQGra..32e5008S} investigated the dynamics of titled Bianchi V models in the presence of diffusion. van den Hoogen and Coley \cite{1995CQGra..12.2335V} performed a qualitative analysis of Bianchi Type V cosmological models with a viscous fluid. Billyard, et.al \cite{1999CQGra..16.4035B} investigated Bianchi Type V models with scalar fields and barotropic matter. Belinch{\'o}n \cite{2012Ap&SS.338..381B} studied the dynamics of different physical constants in the context of a Bianchi Type V model. 

It is of our interest to see if starting from a more general spacetime such as the Bianchi Type V, whether an AdS spacetime is an asymptotic state in the phase space of the dynamics of this model. Further, because of the arbitrary nature of the cosmological constant that we are considering, do other interesting behaviours occur as part of these dynamics as well? Such questions to the best of the author's knowledge have not been given in the literature till date. Further, an analysis of the dynamics of a vacuum Bianchi Type V model with an arbitrary cosmological constant has not been given either. We employ a dynamical systems approach throughout. This work will hopefully shed light on both topics. 

\section{The Einstein Field Equations}
In the section, we derive the equations that describe the dynamics of our vacuum Bianchi Type V model with a cosmological constant as a free parameter. In doing so, we make use of the orthonormal frame approach of Ellis and MacCallum \cite{ellismac} \cite{hervik}. In this approach, the Einstein field equations for the non-titled Bianchi models take the form of two curvature propagation equations:
\begin{equation}
\label{eq:adot}
\dot{a}_{i} + \frac{1}{3}\theta a_i  + \sigma_{ij}a^{j} + \epsilon_{ijk}a^{j} \Omega^{k} = 0,
\end{equation}
and
\begin{equation}
\label{eq:ndot}
\dot{n}_{ab} + \frac{1}{3}\theta n_{ab} + 2 n^{k}_{(a}\epsilon_{b)kl}\Omega^{l} - 2 n_{k (a} \sigma_{b})^{k} = 0,
\end{equation}
a shear propagation equation:
\begin{equation}
\label{eq:sigdot}
\dot{\sigma}_{ab} + \theta \sigma_{ab} - 2\sigma^{d}_{ (a}\epsilon_{b)cd}\Omega^{c} + ^{(3)}R_{ab} - \frac{1}{3}h_{ab} ^{(3)}R = \pi_{ab},
\end{equation}
Raychaudhuri's equation:
\begin{equation}
\label{eq:raych1}
\dot{\theta} + \frac{1}{3}\theta^2 + \sigma_{ab}\sigma^{ab} + \frac{1}{2} \left(\mu + 3p\right) - \Lambda = 0,
\end{equation}
and the Friedmann equation:
\begin{equation}
\label{eq:fried1}
\frac{1}{3}\theta^2 = \frac{1}{2}\sigma_{ab}\sigma^{ab} - \frac{1}{2} ^{(3)}R + \mu + \Lambda.
\end{equation}
In these equations, the Ricci 3-tensor $^{(3)}R_{ab}$ is given by
\begin{eqnarray}
^{(3)}R_{ab} = -2 \epsilon^{cd}_{(a}n_{b)c}a_d + 2 n_{ad}n^{d}_{b} &-& \nonumber \\ n n_{ab} - h_{ab} \left(2 a^2 + n_{cd}n^{cd} - \frac{1}{2}n^2\right),
\end{eqnarray}
which implies that the Ricci 3-scalar $^{(3)}R$ is given by
\begin{equation}
\label{eq:ricciscalar}
^{(3)}R = - \left(6a^2 + n_{cd}n^{cd} - \frac{1}{2}n^2\right).
\end{equation}
Further, $h_{ab}$ denotes the spatial metric of the orthonormal frame, which is simply $\text{diag}(1,1,1)$, and $\pi_{ab}$ denotes the anisotropic stress terms in the fluid, which we take to be zero since we are considering vacuum models with nonzero $\Lambda$. Further to these equations, there is also a non-trivial constraint equation given by
\begin{equation}
\label{eq:nontriv1}
3a^{b} \sigma_{ba} - \epsilon_{abc} n^{cd} \sigma^{b}_{d} = 0.
\end{equation}

Now, for a Bianchi Type V model, we require that $a^i = (0,0,a)$, where $a > 0$. Further, $n_{ij} = \text{diag}(n_1, n_2, n_3) = (0,0,0)$. With this information, we see that Eq. \eqref{eq:nontriv1} yields
\begin{equation}
\sigma_{31} = \sigma_{32} = \sigma_{33} = 0.
\end{equation}
Further, since the shear tensor $\sigma_{ab}$ is trace-free, i.e., $\sigma^{a}_{a} = 0$, there are only two independent non-zero shear components. Hence,
\begin{equation}
\sigma_{ab} = \text{diag}(\sigma_{+}, -\sigma_{+}, 0).
\end{equation}
From Eq. \eqref{eq:adot}, we have that $\Omega^{1} = \Omega^{2} = 0$. The shear propagation equation \eqref{eq:sigdot} further implies that $\Omega^{3} = 0$. 

Therefore, equations governing the dynamics of our model are
\begin{eqnarray}
\dot{a} &=& -\frac{1}{3}\theta a, \\
\dot{\sigma}_{+} &=& -\theta \sigma_+, \\
\dot{\theta} &=& -\frac{1}{3}\theta^2 - 2\sigma_{+}^2 + \Lambda, \\
\label{eq:fried2}
\frac{1}{3}\theta^2 &=& \sigma_{+}^2 + 2a^2 + \Lambda. 
\end{eqnarray}

To slightly simplify things, we will use the Friedmann equation to eliminate the shear $\sigma_{+}$ from the Raychaudhuri equation, thus yielding a planar dynamical system that describes the dynamics of our Bianchi Type V model:
\begin{eqnarray}
\label{eq:raych2}
\dot{\theta} &=& -\theta^2 + 4 a^2 + 3 \Lambda, \\
\label{eq:adot2}
\dot{a} &=& -\frac{1}{3}\theta a. 
\end{eqnarray}

\section{A Qualitative Analysis}
With the dynamical equations \eqref{eq:raych2}-\eqref{eq:adot2} in hand, we are now in a position to describe the dynamics of the cosmological model under consideration. The dynamical system yields two equilibrium points for $\Lambda < 0$, one equilibrium point for $\Lambda = 0$, and two equilibrium points for $\Lambda > 0$. 

First, for the case of a positive cosmological constant, $\Lambda > 0$. The equilibrium points are
\begin{equation}
P_{1,2}: (\theta^{*}, a^{*}) = \left(\pm \sqrt{3} \sqrt{\Lambda}, 0 \right),
\end{equation}
which represent expanding and contracting de Sitter universes respectively. The eigenvalues for $P_{1}$ are found to be
\begin{equation}
\label{eq:eigs1}
\lambda_{1,2} = -2 \sqrt{3} \sqrt{\Lambda}, \quad -\frac{\sqrt{\Lambda}}{\sqrt{3}}.
\end{equation}
Since for $\Lambda > 0$, $\lambda_{1,2} < 0$, this implies that the expanding de Sitter universe is a stable node of the dynamical system. 

For the contracting de Sitter universe described by $P_{2}$, the eigenvalues are found to be
\begin{equation}
\lambda_{1,2} = 2 \sqrt{3} \sqrt{\Lambda}, \quad \frac{ \sqrt{\Lambda}}{\sqrt{3}}.
\end{equation}
Since for $\Lambda > 0$, $\lambda_{1,2} > 0$, this implies that the contracting de Sitter universe is an unstable node of the dynamical system. 

For the case of a negative cosmological constant, $\Lambda < 0$, there exist two additional equilibrium points. In particular, these are given by
\begin{equation}
P_{3,4}: (\theta^{*}, a^{*}) = \left(0, \pm \frac{1}{2} \sqrt{3} \sqrt{ |\Lambda|}\right).
\end{equation}
In both cases, one sees that the Ricci 3-scalar according to Eq. \eqref{eq:ricciscalar} has the value
\begin{equation}
^{(3)}R = -\frac{9}{2} |\Lambda| < 0.
\end{equation}
Therefore, these equilibrium points represent static Anti-de Sitter universes. For both $P_{3}$ and $P_{4}$, the eigenvalues are found to be
\begin{equation}
\lambda_{1,2} = \pm i \sqrt{2} \sqrt{ |\Lambda|},
\end{equation}
which implies that both of these points are \emph{center} equilibrium points. 

For the case of a zero cosmological constant, $\Lambda = 0$,  there exists a single equilibrium point, namely,
\begin{equation}
P_{5}: (\theta^{*}, a^{*}) = \left(0,0\right),
\end{equation}
which represents a Minkowski spacetime. The eigenvalues at this point are given by
\begin{equation}
\lambda_{1,2} = 0,0.
\end{equation}

In summary, we see that for a positive cosmological constant, there exist one-dimensional stable and unstable manifolds in the neighbourhoods of $P_1$ and $P_2$ respectively, and two-dimensional centre manifolds in the neighbourhoods of $P_{3,4,5}$ respectively. Hence, the stability of these points cannot be determined by linearization methods.

\section{Bifurcations}
The presence of the center manifolds corresponding to $\Lambda < 0$ and $\Lambda = 0$ in the dynamical system indicate that this system exhibits some interesting bifurcation behaviour, which we will attempt to describe in this section. Bifurcations occur through destabilizations of the dynamical system. These can be seen as follows. Consider the linearization of Eqs. \eqref{eq:raych2}-\eqref{eq:adot2} in the neighbourhood of $P_{1}$. We have that
\begin{eqnarray}
\dot{\theta} &=& -2 \sqrt{3}\sqrt{\Lambda} \\
\dot{a} &=& -\frac{\sqrt{\Lambda}}{\sqrt{3}}.
\end{eqnarray}
Therefore, $P_{1}$ is destabilized by $\theta$ and $a$ for $\Lambda = 0$. 

The linearization of Eqs. \eqref{eq:raych2}-\eqref{eq:adot2} in the neighbourhood of $P_{2}$ yields
\begin{eqnarray}
\dot{\theta} &=& 2 \sqrt{3}\sqrt{\Lambda} \\
\dot{a} &=& \frac{\sqrt{\Lambda}}{\sqrt{3}}.
\end{eqnarray}
Therefore, $P_{2}$ is destabilized by $\theta$ and $a$ for $\Lambda = 0$.

The linearization of Eqs. \eqref{eq:raych2}-\eqref{eq:adot2} in the neighbourhood of $P_{3}$ yields
\begin{eqnarray}
\dot{\theta} &=& 4 \sqrt{3} \sqrt{|\Lambda|} a \\
\dot{a} &=& -\frac{ \sqrt{|\Lambda|}}{2 \sqrt{3}} \theta.
\end{eqnarray}
Therefore, $P_{3}$ is destabilized by $\theta$ and $a$ for $\Lambda = 0$.

The linearization of Eqs. \eqref{eq:raych2}-\eqref{eq:adot2} in the neighbourhood of $P_{4}$ yields
\begin{eqnarray}
\dot{\theta} &=& -4 \sqrt{3} \sqrt{|\Lambda|} a \\
\dot{a} &=& \frac{ \sqrt{|\Lambda|}}{2 \sqrt{3}} \theta.
\end{eqnarray}
Therefore, $P_{3}$ is destabilized by $\theta$ and $a$ for $\Lambda = 0$.

\begin{figure}[h]
\centering
\includegraphics[scale=0.70]{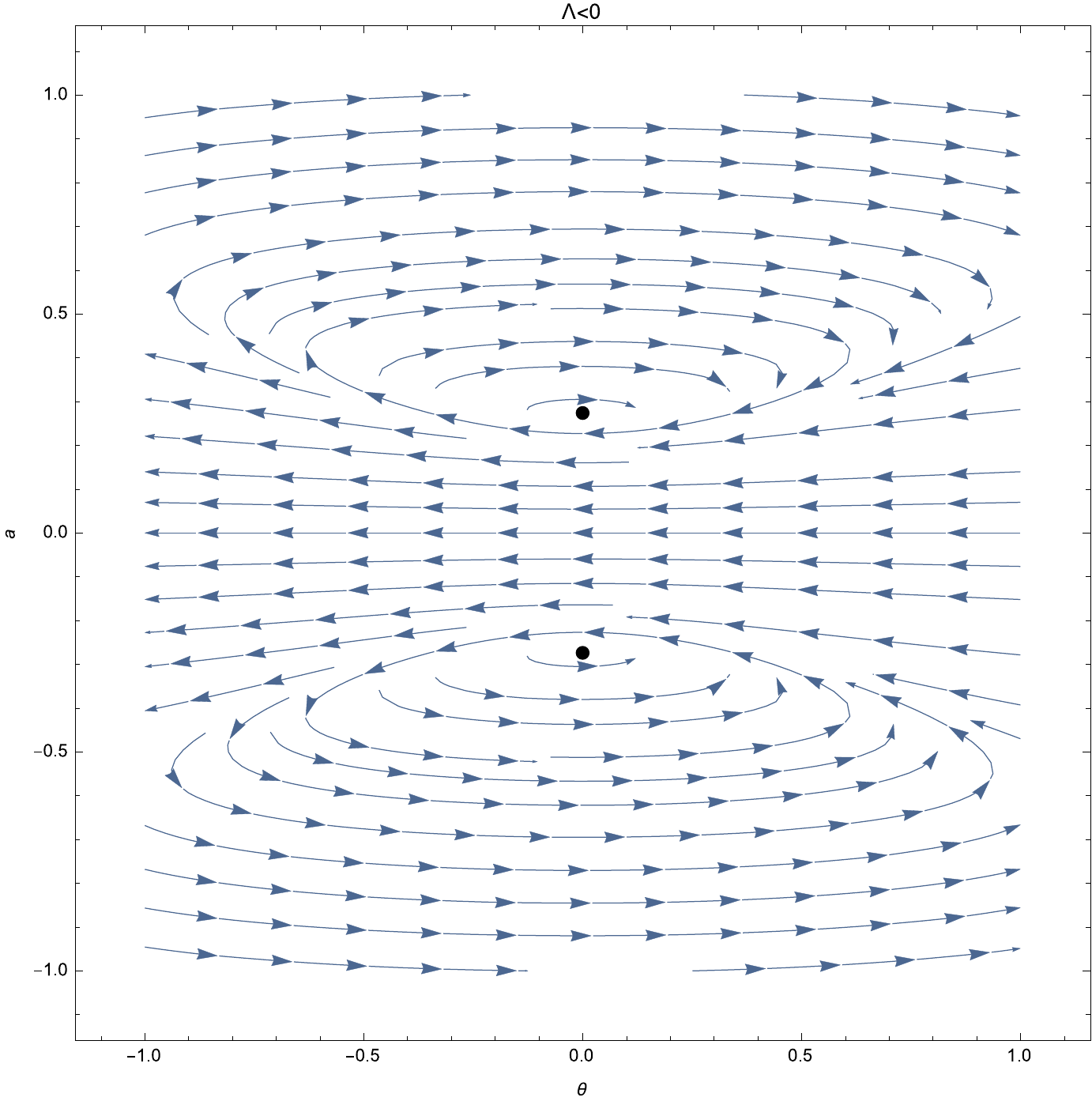}
\caption{A phase plot of the dynamical system for $\Lambda < 0$. The two AdS equilibrium points $P_{3,4}$ are indicated by dots. One can clearly see the formation of a periodic solution.}
\label{fig:fig1}
\end{figure}

\begin{figure}[h]
\centering
\includegraphics[scale=0.70]{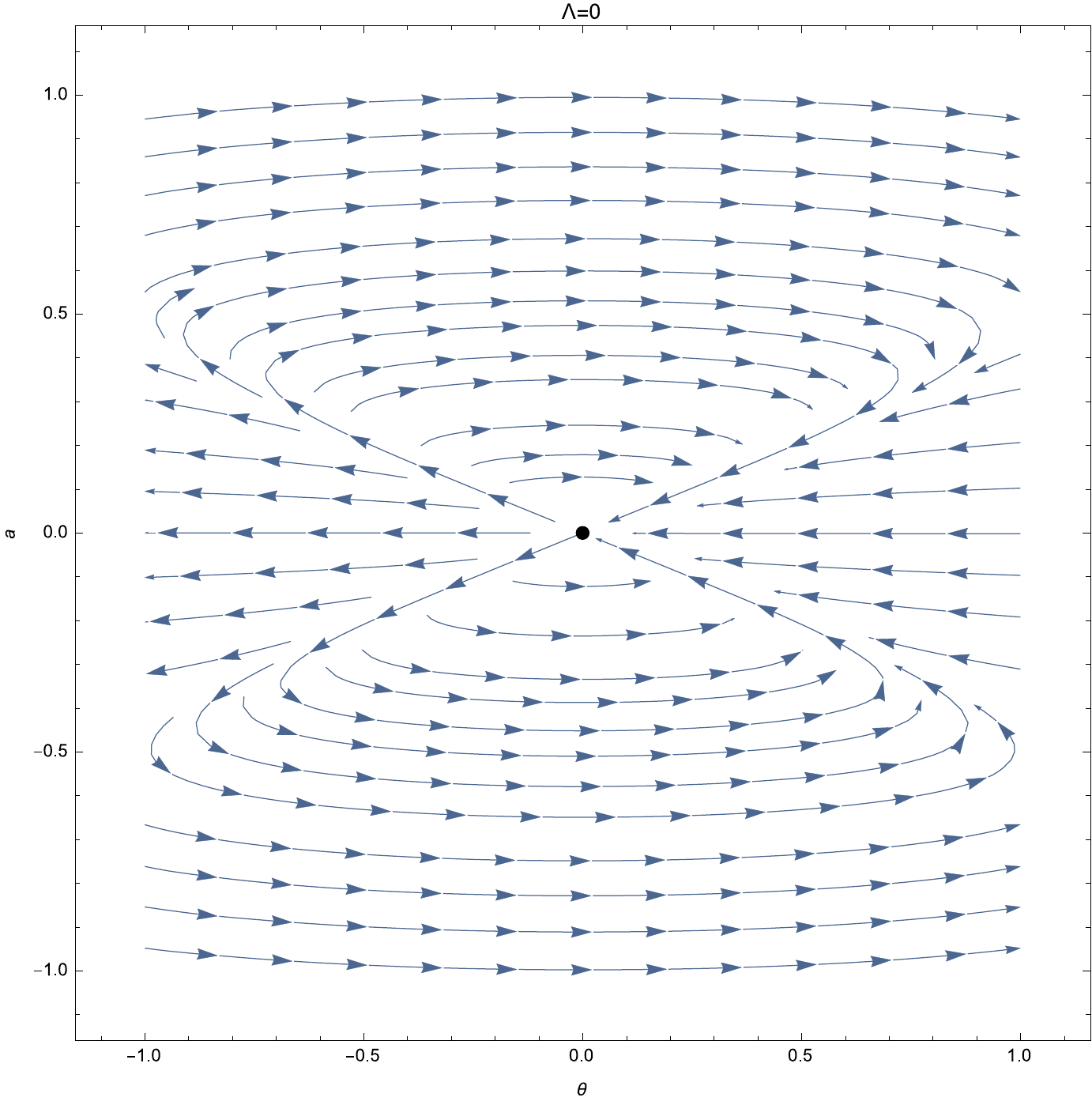}
\caption{A phase plot of the dynamical system for $\Lambda = 0$. The Minkowski equilibrium point $P_{5}$ is indicated by a dot. One can clearly see the formation of a homoclinic orbit connecting this point to itself.}
\label{fig:fig2}
\end{figure}

\begin{figure}[h]
\centering
\includegraphics[scale=0.65]{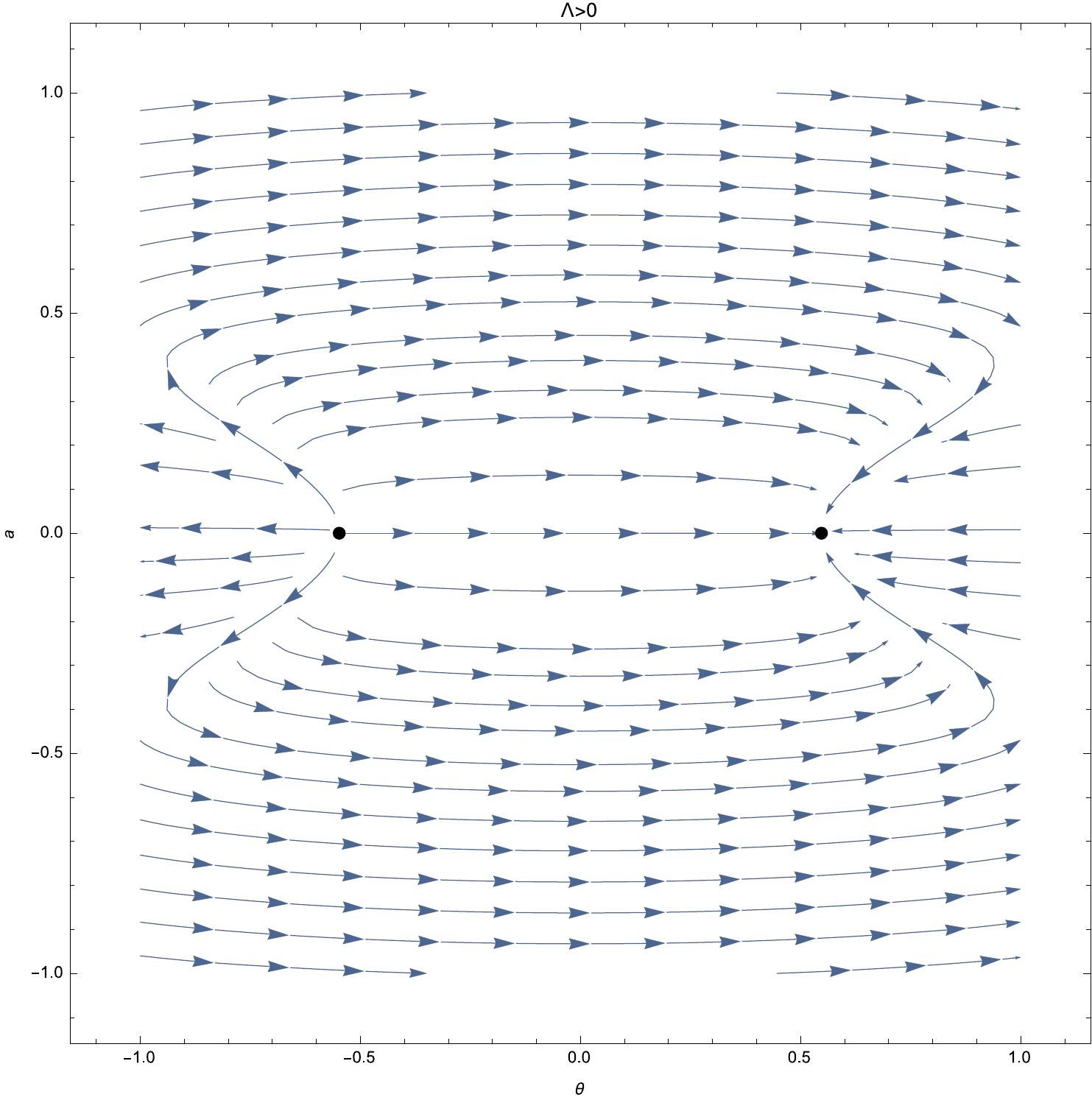}
\caption{A phase plot of the dynamical system for $\Lambda > 0$. The two de Sitter equilibrium points $P_{1,2}$ are indicated by dots. One can clearly see the stable and unstable node behaviour as indicated from the stability analysis of the equilibrium points. There is also clearly a heteroclinic orbit that connects the expanding and contracting de Sitter equilibrium points.}
\label{fig:fig3}
\end{figure}

We therefore see that the line $\Lambda = 0$ in the parameter space dictates the bifurcations of the dynamical system. In Figs. \ref{fig:fig1}-\ref{fig:fig3}, we present some phase portraits of the system when $\Lambda < 0$, $\Lambda = 0$, and $\Lambda > 0$ clearly showing this bifurcation behaviour of the dynamical system. 

In fact, one can see that as $\Lambda$ becomes negative, the system begins to show periodic behaviour. Indeed, the eigenvalues associated with $P_{3,4}$ are both complex with no real parts. This indicates that the type of bifurcation that occurs in this system is a degenerate Andronov-Hopf bifurcation \cite{kuznet} when $\Lambda < 0$. Further, when $\Lambda = 0$, one sees that the oscillatory behaviour persists but dissipates when $\Lambda > 0$. Further, there is a homoclinic orbit present for $\Lambda = 0$ that connects the Minkowski equilibrium point to itself. When $\Lambda > 0$, a heteroclinic orbit forms connecting the expanding and contracting de Sitter equilibrium points.

\section{Further Results}

Consider the case where $\Lambda > 0$, we will try to ascertain some information about the asymptotic stability of $P_{1}$, the expanding de Sitter universe. We make use of the following result \cite{arnolddyn}: If all eigenvalues of the linear part of a vector field at a singular point have negative real part, then the singular point  is asymptotically stable. This result is known is Lyapunov's Theorem on Stability by Linearization. Clearly, looking at the eigenvalues corresponding to $P_{1}$ in Eq. \eqref{eq:eigs1}, one sees that for $\Lambda > 0$, indeed $\lambda_{1,2} < 0$ always. Therefore, indeed, $P_{1}$ is asymptotically stable.

Further, corresponding to $P_{3}$ and $P_{4}$, the eigenvalues found above were both purely imaginary with zero real part, hence, were center equilibrium points. Centers are stable, but not asymptotically stable, since, the orbits will just from closed loops around the equilibrium points, but never converge to the equilibrium points. This can be seen as follows. Following \cite{webb1}, we define a relative entropy function as:
\begin{equation}
\label{eq:relent1}
V(\mathbf{x}) = -\frac{1}{2} \sqrt{3} \sqrt{\Lambda } \log \left(\frac{2 a}{\sqrt{3} \sqrt{\Lambda }}\right).
\end{equation}
Then, the time derivative of $V$ along solution trajectories is found to be
\begin{equation}
\frac{dV}{dt} = -\frac{\sqrt{3} \sqrt{\Lambda } \dot{a}}{2 a}.
\end{equation}
Substituting Eq. \eqref{eq:adot2} into this expression, we finally obtain that
\begin{equation}
\frac{dV}{dt} = \frac{\sqrt{\Lambda } \theta}{2 \sqrt{3}}.
\end{equation}
Now, at $P_{3}$ and $P_{4}$, we have that $\theta = 0$. Therefore, $dV/dt = 0$. Since the time derivative of the relative entropy function is zero at both $P_{3}$ and $P_{4}$, solutions are indeed periodic about these points.

%

\section{Conclusions}
In this paper, we described the dynamics of a Bianchi Type V vacuum universe with an arbitrary cosmological constant that behaved as a parameter in the resulting dynamical system. We began by using an orthonormal frame approach to write Einstein's field equations as a coupled system of first-order ordinary differential equations. We then computed the equilibrium points which were found to be expanding and contracting de Sitter universe solutions, a Minkowski spacetime solution, and static Anti-de Sitter universe solutions, which were characterized by a negative cosmological constant in addition to constant negative spatial curvature. While the expanding de Sitter universe solution was found to be asymptotically stable for $\Lambda > 0$, we also found that the AdS solution was stable for $\Lambda < 0$. Further, since by the local stability analysis we found that this solution behaved as a center equilibrium point, this implied that for $\Lambda < 0$, one has a cosmological model which has oscillatory behaviour. We further described the bifurcation behaviour of the system in terms of $\Lambda$.

\section{Acknowledgements}
The author would like to thank George F.R. Ellis for interesting discussions pertaining to AdS universes. The author would also like to thank Michael C. Haslam for interesting discussions regarding dynamical systems.

\newpage 
\bibliographystyle{ieeetr} 
\bibliography{sources}%

\end{document}